\documentclass[iop]{emulateapj}

\slugcomment{draft version}

\shorttitle{Observed Abundances and Relative Frequencies of Amino Acids in Meteorites}
\shortauthors{Cobb \& Pudritz}

\usepackage{tabularx}
\usepackage{colortbl}
\usepackage{rotating}
\usepackage{setspace}
\usepackage{etex}
\usepackage{m-pictex}
\usepackage{m-ch-en}
\usepackage[toc,page]{appendix}

\begin{document}

\title{NATURE'S STARSHIPS. I. OBSERVED ABUNDANCES AND RELATIVE FREQUENCIES OF AMINO ACIDS IN METEORITES}

\author{Alyssa K. Cobb\altaffilmark{1,2,3} and Ralph E. Pudritz\altaffilmark{1,2,4}}

\altaffiltext{1}{Origins Institute, McMaster University, ABB 241, 1280 Main Street,  Hamilton, ON, L8S 4M1, Canada}
\altaffiltext{2}{Department of Physics and Astronomy, McMaster University, ABB 241, 1280 Main Street, Hamilton, ON, L8S 4M1, Canada}
\altaffiltext{3}{E-mail: cobbak@mcmaster.ca}
\altaffiltext{4}{E-mail: pudritz@physics.mcmaster.ca}

\begin{abstract}
The class of meteorites called carbonaceous chondrites are examples of material from the solar system which have been relatively unchanged from the time of their initial formation. These meteorites have been classified according to the temperatures and physical conditions of their parent planetesimals. We collate available data on amino acid abundance in these meteorites and plot the concentrations of different amino acids for each meteorite within various meteorite subclasses. We plot average concentrations for various amino acids across meteorites separated by subclass and petrologic type. We see a predominance in the abundance and variety of amino acids in CM2 and CR2 meteorites. The range in temperature corresponding to these subclasses indicates high degrees of aqueous alteration, suggesting aqueous synthesis of amino acids. Within the CM2 and CR2 subclasses, we identify trends in relative frequencies of amino acids to investigate how common amino acids are as a function of their chemical complexity. These two trends (total abundance and relative frequencies) can be used to constrain formation parameters of amino acids within planetesimals. Our organization of the data supports an onion shell model for the temperature structure of planetesimals. The least altered meteorites (type 3) and their amino acids originated near cooler surface regions. The most active amino acid synthesis likely took place at intermediate depths (type 2). The most altered materials (type 1) originated furthest toward parent body cores. This region is likely too hot to either favor amino acid synthesis or for amino acids to be retained after synthesis.


\end{abstract}

\section{Introduction}

Meteorites provide a geological and chemical record of the formation of organic molecules in the early solar system. Meteoritic parent planetesimals formed during the formation of the solar system and contain materials present at that time. Likely composed primarily of rock and ice, these parent bodies would be on the order of tens of kilometers in diameter. The closest analogues we have for these bodies are present-day asteroids. Meteorites found on Earth are likely fragments of their larger parent bodies, ejecta from collisions with parent bodies in the protoplanetary disk \citep{Sephton2005,Gaidos2007}. Carbonaceous chondrites are thought to originate from type C asteroids, which are water-rich and likely came from the outer regions of the asteroid belt \citep[and references therein]{Morbidelli2012}. There are a wide variety of meteorite types, all composed of a wide range of varying materials. Being inhomogeneous in nature, different compositions of meteorite are thought to indicate the degree of mixing of materials when their parent bodies were formed. 

Amino acids that formed in meteoritic parent planetesimals are an important source of organic material that was delivered to young planets via meteoritic impacts. They are the basic building blocks in proteins used by life on Earth. This extraterrestrial origin of biomolecules suggests one possible way that the prebiotic conditions for life in the universe could have been established, both on our own early Earth and perhaps on other planets \citep{Chyba1992,BottaBada,Sephton2002,Glavin2011}.

In this paper we investigate patterns in both total abundance and relative frequencies of amino acids in carbonaceous chondrites. Identification of these patterns offer insights into the physics behind their formation in planetesimals. Given the long time frame that parent bodies are likely to be in an aqueous state (a few million years \citep{Travis2005}), it is possible to apply principles of equilibrium thermodynamics to follow the synthesis of amino acids. Total amino acid abundances depend on both temperature and the initial concentrations of reactants involved in the chemical processes that create them. Total abundance patterns therefore act as tracers of both temperature and molecular abundances within a parent body. In Paper II of this series (A. K. Cobb \& R. E. Pudritz 2014, in preparation), we use thermodynamic principles and Gibbs free energies to model the synthesis of amino acids in a theoretical planetesimal.

Given some basic understanding of the formation of meteorites, it is now feasible to model meteoritic parent bodies in an effort to constrain their internal chemistry. \citet{Young1999}, \citet{Schulte2004} and \citet{Travis2005} model parent bodies as agglomerations of ice, rock, and various interstellar organic material present during formation. The radionuclide decay of isotopes such as $^{26}$Al in their interiors keeps the temperature near the core high enough to melt water. Once liquid water is present, there are a variety of internal chemical processes that cause the organic material initially present to alter into new substances. These processes are collectively called aqueous alteration. See \citet[and references therein]{Grimm1989,McSween1989,Weiss2013} for a review.

The class of meteorites called carbonaceous chondrites is of particular interest with regard to the origin of organic material. Representing approximately 4\% of meteorite falls \citep{Glavin2011}, these meteorites are relatively undifferentiated from the original material with which they originally formed. They contain high concentrations of water, carbon, and various organic materials. The carbonaceous chondrites are divided into subclasses and petrologic types based on their composition, mineralogy, and degrees of internal alteration. There is a range in alteration temperatures associated with different minerals found in meteorites, and these alteration temperatures are used to categorize meteorite samples. There are two types of alteration processes, aqueous and thermal. Carbonaceous chondrites that have undergone aqueous alteration show increased levels of organic material, including amino acids \citep{Sephton2002,Weisberg2006,Glavin2011}.

Meteorite classification is based on two characteristics in particular: primarily, chemical composition, and secondarily, the amounts of aqueous and thermal processing. Aqueous alteration is a general term for the transformation of molecules to organic material in a hydrated environment - exactly what is occurring in the interior of meteoritic parent bodies resulting in the formation of amino acids. Thermal metamorphism occurs when alteration temperatures are too hot to retain a hydrated environment. See \citet{Kojima1996}, \citet{Weisberg2006} and \citet{Weiss2013} for a review.

The range of temperatures that allows for significant amounts of this aqueous alteration is roughly $0^{\circ}$C - $300 ^\circ$C. At relatively high temperatures, $>300^\circ$C, the parent bodies undergo thermal metamorphism instead; no aqueous alteration occurs, and we see little to no synthesis of organics. Petrologically speaking, more aqueously altered meteorites are given a classification value of type 1 or type 2, and thermally altered meteorites are listed as type 4 - 6. Type 3 meteorites are the most pristine samples of early solar system material, having experienced minimal degrees of either aqueous alteration or thermal metamorphism \citep[and references therein]{Sephton2002,Weisberg2006}.

There is great diversity amongst carbonaceous chondrite literature regarding petrographic types and their corresponding alteration temperatures. These are well summarized in \citet{Grimm1989}, \citet{Morlok2013} and \citet[and references therein]{Weiss2013}. See Section \ref{classification} for a discussion of alteration temperature classification boundaries among petrographic types as well as the physical differences among meteorite classifications. Broadly speaking, the different petrographic types are classified based on compositional differences of meteorite samples. In each meteorite, minerals have undergone different degrees of melting or other alteration process, and each mineral has a unique melting point. Based on which minerals have been differentiated, a meteorite may be classified as having undergone alteration temperatures that reach the melting point of those specific minerals.


We explore the abundance patterns of $\alpha$-proteinogenic amino acids found in carbonaceous chondrites. These are among the most common amino acids and are among the 20 used in our genetic code. We collate a large quantity of available data on amino acid concentration in a variety of carbonaceous chondrites from numerous sources and different laboratories. The data are shown graphically to more easily distinguish patterns in amino acid abundance across meteorite subclasses. In particular we use the data to identify patterns in amino acid concentration per meteorite class and per amino acid, as well as a total amino acid concentration per meteorite. We average total amino acid concentrations per meteorite class to observe overall abundance trends across the carbonaceous chondrites.

In addition to identifying abundance patterns across subclasses, we also investigate the relative frequencies of amino acids within two subclasses (CM and CR) of carbonaceous chondrites. \citet{Higgs2009} examined a very limited set of carbonaceous chondrite data and found an empirical relation between the frequencies of some amino acids and their Gibbs free energies. The order they found related to the most common amino acids they called 'early-type', referring to the set of amino acids likely present on early Earth. These amino acids are glycine, alanine, aspartic acid, glutamic acid, valine, serine, isoleucine, leucine, proline, and threonine. Perhaps because of the availability of these amino acids due to their energetic favorability, they were incorporated into our genetic code. Increasing molecular complexity of amino acids is reflected in the increasing values of their associated Gibbs energies of formation. We expand on the amino acid landscape explored in \citet{Higgs2009}.

Our treatment of the data is designed to emphasize patterns of amino acid synthesis in planetesimals across a range of alteration temperatures spanning meteorite subclasses. We also show the relative abundance patterns of amino acids in various meteorites based on environmental factors and internal chemistry during synthesis. Our analysis of these data shows that the CM and CR type 2 meteorites occupy a 'sweet spot' in alteration, which corresponds to specific meteorite classifications at which amino acid abundances peak. The patterns that emerge from the cumulative data plots reveal an optimal petrographic value of type 2, generally associated with temperatures between $0^{\circ}$C and $100^{\circ}$C. These values correspond to increased levels of aqueous alteration and the synthesis of amino acids. From these emerging patterns in abundance and frequency, we infer information about the chemical composition and temperatures at which amino acids would have likely been synthesized in parent planetesimals. The data collated here help provide insights into the chemistry behind the formation of these amino acids in parent bodies.

\section{Amino Acids}

There are hundreds of amino acids found on Earth. In this work, we emphasize a specific subset: $\alpha$-proteinogenic amino acids, so named for the location of their amino groups (NH$_2$). The general structure of an amino acid is 
\startchemical[height=2900,width=4000]
\hspace{-6mm}
\chemical[ONE,Z01357,SB1357][C_{\alpha},COOH,H,R,NH_2] 
\stopchemical,
with the R group indicating the side chain. It is this side chain that varies and creates the different amino acids. The '$\alpha$' signifies that the amino group of the amino acid is attached at the $\alpha$ carbon. The $\alpha$ carbon is the first carbon; the carbon to which the carboxyl group (COOH) and side chain (R) attach (Pizzarello 2009; Pizzarello \& Holmes 2009; J. Hein 2012, private communication). 
Note that the Strecker synthesis processes discussed below create only $\alpha$-amino acids.

Amongst the considerable variety of $\alpha$-amino acids, we consider only those amino acids that are proteinogenic. Proteinogenic amino acids are a set of 20 amino acids coded for directly by our genetic code. They are an essential part of life on Earth, as they are the building blocks and precursors to proteins. These 20 amino acids are: glycine, alanine, aspartic acid, glutamic acid, valine, serine, isoleucine, leucine, proline, threonine, lysine, phenylalanine, arginine, histidine, asparagine, glutamine, cysteine, tyrosine, methionine, and tryptophan. As in \citet{Higgs2009}, they may be divided into early-type amino acids (glycine through threonine) and late-type amino acids (lysine through tryptophan).

\subsection{Strecker Synthesis}

As mentioned earlier, once liquid water is present in the meteoritic parent body, it and any organic materials present may undergo aqueous alteration to produce amino acids. In our work, we focus on one particular type of aqueous alteration, a process called Strecker synthesis. We refer to a synthesis pathway defined as reactions between aldehydes with hydrogen cyanide, ammonia, and water to create amino acids \citep{Miller1957,Peltzer1984,Schulte1995,Botta2002,Sephton2005,Elsila2007}. Strecker theory is a promising avenue of investigation when considering the synthesis of extraterrestrial amino acids; all requisite reactants exist among the volatile content of comets, which we use as analogues for material that would also have been incorporated into meteoritic parent bodies \citep{Schulte2004,Alexander2011}. Early work on Strecker synthesis was completed by \citet{Miller1957} in his electric discharge experiments, and was followed by \citet{Peltzer1984}, who worked on the chemical reactions leading to amino acids and the effects of ammonia concentration on the pathway.

We restrict our work to Strecker synthesis reactions defined in the following manner: aldehyde molecules, in aqueous solution, with hydrogen cyanide and ammonia, react to produce $\alpha$-amino acids. The initial reactants in Strecker processes combine to create various intermediaries, which themselves undergo a one-way process called hydrolysis, resulting in the formation of $\alpha$-amino acids. The simplest example of a Strecker pathway involves glycine, the simplest amino acid, shown in Equation (1).
\vspace{2mm}
%
%

 $$\hspace{-4mm}
 \setupchemical[width=fit,size=small]
    \chemical{CH_2O}{~formaldehyde} \chemical{PLUS} \chemical{HCN}{hydrogen~cyanide} \chemical{PLUS} \chemical{H_2O}{water} \chemical{PLUS} \chemical{NH_3}{ammonia}  
 $$
 
 $$
 \setupchemical[width=fit,size=small]  \chemical{GIVES} \chemical{C_2H_5NO_2}{glycine} \chemical{PLUS} \chemical{NH_3}{ammonia}     \hspace{25mm} (1)$$ 

In the above pathway, we see a mass balance Strecker reaction for formaldehyde (also known as methanal) reacting with hydrogen cyanide, water, and ammonia. The products of this reaction are the amino acid glycine and another ammonia molecule. 
It is important to note that even though one ammonia molecule is destroyed initially and one is created at the end, it is chemically inaccurate to depict this reaction as occurring without ammonia altogether. The initial nitrogen, contributed by ammonia, remains in the amino acid. It is the nitrogen contributed by the HCN molecule that later breaks off from the amino acid and produces the new ammonia molecule. 

In this paper, we restrict our attention to concentrations of $\alpha$-amino acids that may have been produced by Strecker reactions. There are additional possible synthesis pathways which we do not consider. Michael addition, for example, may be responsible for the formation of $\beta$-amino acids in parent bodies, which have their characteristic amino groups attached at the $\beta$-carbon, immediately adjacent to the $\alpha$-carbon \citep{Ehrenfreund2001,Martins2007,Glavin2009}. We also do not consider alternate Strecker pathways, in which the reactants are ketones. For example, a Strecker reaction taking place with a simple ketone rather than formaldehyde, and also reacting with no ammonia present, creates glycolic acid rather than glycine \citep{Peltzer1984,Schulte1995,Ehrenfreund2001}. 

\citet{Glavin2006}, \citet{Glavin2011} and \citet{Martins2007} detected $\beta$-alanine and $\gamma$-aminoisobutyric acid during analyses of carbonaceous chondrites. Murchison meteorite and its impressive array of amino acids also includes  $\beta$-amino acids, $\gamma$-amino acids, and non-proteinogenic $\alpha$-amino acids, e.g., $\beta$-alanine, $\gamma$-amino-n-butyric acid, aminoisobutyric acid, norvaline, and isovaline \citep{Sephton2002}. Laboratory analyses of retrieved meteorite falls have revealed the presence of a significant number of amino acids and other organic material, such as aldehydes and carboxylic acids, within meteorites. There is a diverse assortment of amino acids that has been found in meteorites. \citet{Burton2012b} present an up-to-date and cumulative list of amino acids discovered in meteorites thus far.

\subsection{Possibility of Contamination}

As subunits of Earth proteins, proteinogenic amino acids are important for origins of life research. However, the proteinogenic amino acid content in meteorites is therefore also likely to be affected by terrestrial contamination. Distinguishing between abiotically produced amino acids that originated in a parent body and amino acids introduced to a meteorite following its arrival on Earth is a challenging task. Work by \citet{Botta2002}, \citet{Glavin2009}, \citet{Glavin2011}, and \citet{Burton2012a}, for example, discuss laboratory methods to distinguish between terrestrial and extraterrestrial amino acids. 

Most amino acids come in one of two configurations, D- and L-. Amino acids created abiotically produce ratios of D:L molecules $\sim 1$, called a racemic mixture. Proteins on Earth, however, are made with L amino acids. The ratio of D:L enantiomers is nonracemic. The greater abundance of L amino acids than D amino acids results in a low D:L, $< 1$. Measured D:L ratios of amino acids in meteorites are used to indicate the degree, if any, of terrestrial contamination. A low D:L ratio indicates the amino acid content of a meteorite has been affected by terrestrial contamination. Racemic levels of D:L indicate the amino acid content of chondrite has not been affected by terrestrial contamination. A racemic mixture suggests an amino acid content likely not synthesized on Earth. The ratio must have been synthesized racemic prior to the meteorite's arrival on Earth. 

Samples of meteorites Murray, Murchison, and Mighei, for example, likely contain amino acid contents affected by varying degrees of terrestrial contamination \citep{Botta2002,Glavin2011}. However, meteorite samples such as Murchison USNM 6650, LEW 90500, LON 94102, EET 92042, and QUE 99177 all contain D:L ratios $\sim 1$, suggesting abiotic origins and not terrestrial contamination \citep{Glavin2011}. 

We make no attempt to further correct for reported amino acid abundances that may have been affected by terrestrial contamination beyond what may have been attempted during laboratory analyses. Full considerations of contamination possibilities and abundance level corrections can be found in the original sources listed for each chondrite. 

\section{Meteoritic Classification}\label{classification}

In this section we briefly review the basics of meteorite classification with the view of emphasizing physical information about the conditions in which they formed. This places constraints on the conditions of amino acid synthesis in their parent bodies.

Meteorites are broadly split into one of three categories based on a coarse initial description: iron, stone, and stony-iron. Stony meteorites are further divided into one of two categories: chondrites and achondrites. The chondrites are the most primitive examples of meteorites. They are relatively undifferentiated from the makeup of their parent body material and their chemical composition is of near-solar abundance. These meteorites reflect processes occurring during and following the formation of the solar system. Historically, within the chondrite division are three classes: carbonaceous, ordinary, and enstatite. These terms, while slightly misleading (e.g., 'ordinary' stemming from the fact that these are the most common type), are still used to describe similarities amongst meteorite groups. 

Carbonaceous chondrites have a wide variety of properties that help classify them. Some contain more water, while others may have a greater organic carbon content. The structure of a carbonaceous chondrite may be generalized to an extent; all but a single subclass contain structures called chondrules. Chondrules are near-spherical objects on the order of a few millimeters in size. It is these objects that are examples of molten material residue from the time of the formation of the solar system. Over time, aqueous alteration (or thermal metamorphism, depending on the temperature and water content) took place and led to the creation of organic material \citep{Hayes1967,BottaBada}.

The carbonaceous chondrites are so named for their $>5\%$ by weight composition of organic carbon, which results in a predominantly dark appearance \citep{BottaBada,Botta2005,Glavin2011}. These meteorites collectively represent some of the most primitive samples of meteorite available. Carbonaceous chondrites may be further divided into subclasses, including CI, CM, CR, CH, CO, CV, CK, and CB, based on bulk mineralogical composition. Subclass titles are designated based on the name of a representative fall within each group: CI (Ivuna-like), CM (Mighei-like), CR (Renazzo-like), CH (Allan Hills-like), CO (Ornans-like), CV (Vigarano-like), CK (Karoonda-like), and CB (Bencubbin-like). Varying mineralogical compositions lead to subclass designation, as laid out in \citet{Weisberg2006}. 

CI-type meteorites, like most carbonaceous chondrites, contain near solar abundances of various metals, including magnesium, silicon, and calcium. CI meteorites are unique among carbonaceous chondrites in that they are the only subclass in which no (or almost no) chondrules are found. They have a high content of both water and carbon, both in the form of carbonates and organic carbon, such as amino acids. Their carbon content may be as great as 3\% by weight. The water and carbon content likely indicate that these meteorites have undergone the highest degrees of aqueous alteration. Their parent bodies may have been located near the edges of the protoplanetary disk. They are the most primitive examples of carbonaceous chondrite. \citep{Hayes1967,BottaBada,Weisberg2006}. 

Carbonaceous chondrites of the CM subclass are distinguished by the presence of numerous small chondrules, most no larger than 0.3 mm. The remainder of a CM meteorite is composed of a fine-grained silicate matrix, up to 70\% by volume. Like CI-types, they also contain a variety of metals in near-solar abundances, and contain considerable quantities of nickel. They are very abundant in hydrated minerals, and contain 1.8\% to 2.6\% by weight of carbon. CM chondrites are primarily petrologic type 2, though a few type 1 CM chondrites have also been identified. They show slightly less water content than the CI-types, and, along with the CR subclass, contain the greatest concentrations of organic content of all subclasses \citep{Hayes1967,BottaBada,Weisberg2006}.

CR-type meteorites contain numerous, large chondrules, on the order of 1 mm in size, which occupy approximately 50\% of the available volume. The remaining 50\% is composed of a fine-grain matrix that may be well hydrated. They have the second highest levels of metals in all carbonaceous chondrite subclasses, between 5\% and 8\% by volume. CR meteorites contain a high level of organic material. \citep{Hayes1967,BottaBada,Weisberg2006}. 

CV chondrites contain large chondrules--up to several millimeters in size--and are composed of up to 40\% by volume of a silicate matrix. These meteorites are characterized by their numerous and easily recognizable large chondrules and show minimal evidence of aqueous alteration or thermal metamorphism; they contain relatively little water and organic material. This subclass also contains calcium-aluminum inclusions (CAIs) and approximately 0.2\% to 1.0\% carbon by weight \citep{Hayes1967,BottaBada,Weisberg2006}.

Similar to the CV subclass, CO-type chondrites show minimal evidence for alteration, either thermal or aqueous. They contain few organic molecules. It is thought the CO-types likely formed in the same region of the solar system as the CVs. CO-types also contain CAIs, but they are slightly smaller in size and fewer in number than CAIs in CV meteorites. They have a percent by weight of carbon between 0.2 and 1.0 \citep{Hayes1967,BottaBada,Weisberg2006}.

There are a few additional subclasses of carbonaceous chondrite that we do not consider in this work. The CB (Bencubbin-like) and CK (Karoonda-like) meteorites are both characterized by a lack of internal aqueous alteration. 
CB-types may contain some amino acids, but this classification is recent and there is a limited amount of available amino acid data on CB meteorites (see \citet{Burton2013}). CH chondrites are so named for their extremely high metal content, which may be as high as 40\% \citep{Hayes1967,BottaBada,Weisberg2006}.

Carbonaceous chondrites are additionally categorized into a petrographic type between 1 and 6. They are assigned a subclass and a number, based on their bulk chemical makeup and petrologic type. 
    \begin{figure}[t!]
     \centering
      \includegraphics[width=87mm]{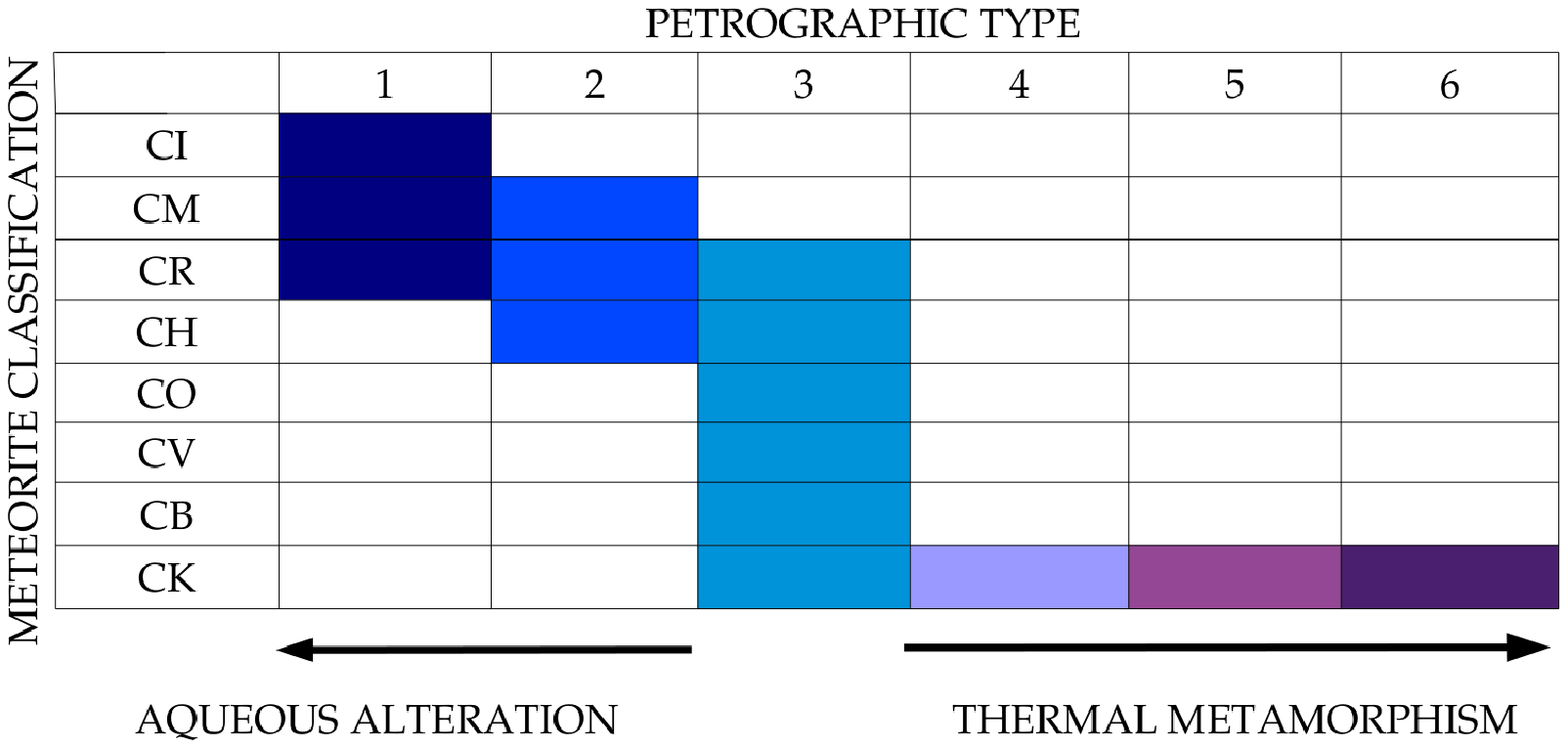}
      \caption{Classification of carbonaceous chondrites. Classified according primarily to chemical composition and secondarily to degrees of thermal and aqueous processing. Adapted from \citet{Sephton2002}. }
      \label{petrology}
    \end{figure}
Figure~\ref{petrology} shows the petrographic types associated with each subclass of carbonaceous chondrite. Type 1 meteorites show the highest degree of aqueous alteration amongst chondrites. 
These meteorites have a high water content and significant organic abundance, including amino acids. They are 3-5\% carbon by weight and 18\% water by weight. Type 2 meteorites are slightly less aqueously altered. 
They contain between 0.8-2.5\% carbon by weight, and between 2-16\% water. Type 3 contains the most pristine meteorite samples. They are in the middle ground, 
showing minimal degrees of either aqueous alteration or thermal metamorphism. By weight, they contain 0.2-1\% carbon and 0.3-3\% water. Type 3 meteorites show an insignificant amino acid abundance corresponding to the limited levels of aqueous processing. Moving from type 4 through type 6 shows an increase in the amount of thermal metamorphism and no aqueous alteration \citep{Sephton2002,Weisberg2006}. 


There is considerable disparity in classification temperatures for different petrographic types. We adopt classification boundaries in generally good agreement with the values most often cited in meteoritic literature. CM meteorites have commonly not experienced alteration temperatures greater than $\sim 50^{\circ}$C. The CI-types, the most aqueously altered, tend to have been heated to a higher temperature than the CM-types. The CR chondrites, which range in petrographic type from 1 through 3, have a correspondingly large range in alteration temperatures possibilities. 

There is fairly good agreement in the literature for the alteration temperature range associated with CI meteorites, all petrographic type 1. \citet{Krot2006}, \citet{Zolensky1993}, and \citet{Huss2006} cite temperatures spanning $50^{\circ}$C to $150^{\circ}$C. \citet{Grimm1989} cite no lower bound, but an upper bound of $150^{\circ}$C.

The CM-types, also, have fairly consistent alteration temperature ranges. \citet{Zolensky1993} cite $0^{\circ}$C to $50^{\circ}$C, and \citet{Huss2006} provide an additional upper bound confirmation of $50^{\circ}$C. \citet{Krot2006} and \citet{Grimm1989} cite a similar range, $0^{\circ}$C to $25^{\circ}$C.

CR chondrites span petrographic types 1, 2, and 3, making temperature classifications for the CR-types relatively broad. \citet{Zolensky1993} cite $50^{\circ}$C to $150^{\circ}$C again, identical to the range cited for CI-types. \citet{Huss2006} provide an upper bound of $150^{\circ}$C. \citet{Morlok2013}, however, give a temperature range of $25^{\circ}$C to $300^{\circ}$C for ``all stages of aqueous alteration'', which we interpret to include types 1 - 3, requiring a hydrated environment.

There is less literature regarding alteration temperatures for CO- and CV-type meteorites, all petrographic type 3. We adopt the value cited in \citet{Zolensky1993}, stating CV-type alteration caps at $50^{\circ}$C. 

These temperature ranges should not be taken as exclusive boundaries, but rather suggestions in range associated with different petrographic types. A picture presents itself when we consider the physical layout of a parent body, supporting the onion-shell model for planetesimals, discussed below.

\subsection{Onion Shell Model for Planetesimals}

It has been proposed that different subclasses of carbonaceous chondrite not only share similar chemical and mineralogical makeups, but may actually share origins. \citet{Ehrenfreund2001} and \citet{Glavin2011}, for example, discuss the possibility that carbonaceous chondrites within a single subclass originate from the same meteoritic parent body. The cause of different subclasses could be that their sources are physically different parent bodies. 

\citet[and references therein]{Weiss2013} discuss the possibility that different petrologic types originate from different layers of a parent body. For example, locations near the core of a parent body might lead to thermal metamorphism (petrologic types 4-6), due to the higher temperatures and limited water content. Among aqueously altered specimens (types 1 - 3), the type 1 meteorites might originate from material nearer to the core of a parent body, and the cooler specimens, types 2 and 3, from regions of a parent body nearer to the surface. The onion shell model for planetesimals is well reviewed in \citet[and references therein]{Weiss2013}, and the fluid flow within such a body is discussed in \citet{Young1999}.

We refer regularly to a temperature classification loosely based on the idea of an onion shell model. We assume the most aqueously altered type 1 meteorites originate from locations within a parent body relatively closer to the core than type 2 and 3 meteorites. Type 3 meteorites, the least altered, are likely to originate from locations nearest to the surface of a planetesimal, associated with comparatively cool temperatures of alteration. Type 2 chondrites, spanning a range in alteration temperatures that encompasses both the higher temperatures of type 1 meteorites and the cooler temperatures of type 3, originate from locations interior to the planetesimal which parallel the temperature classifications. Their material likely resides between layers of type 1 material and type 3 material.

CI meteorites, all type 1, have undergone significant levels of aqueous alteration, but at higher temperatures than their CM and CR fellows. 
CM, and CR-type meteorites, primarily type 2, with a few type 1 and type 3 exceptions, generally contain greater quantities of amino acids and underwent greater degrees of aqueous alteration processing at lower temperatures. CV, CO, and CB-type meteorites are limited to petrologic type 3. These chondrites likely underwent minimal degrees of aqueous alteration at low temperatures. The CK-types, a relatively new subclass, cover a broad range in petrologic types, crossing boundaries between types 3 - 6, moving with increasing formation temperature and levels of thermal processing, up to $950^{\circ}$C \citep{Huss2006}.

The more hydrated environments of the CI/CM/CR meteorites correspond to greater levels of aqueous alteration. Following this classification scheme, we expect to see greater quantities of organic material and amino acids among the aqueously altered meteorites (CI/CM/CR range). However, the most aqueously altered specimens of petrographic type 1 (the CI-types) express lower levels of organics than petrographic type 2 (the CM- and CR-types). \citet{Young1999} and \citet{Weiss2013} suggest possible reasons for this, which include the onion shell model for meteorite origination and different models of fluid flow within chondrite subclasses.  

\section{Meteoritic Amino Acid Data}

The presence of organic constituents in meteoritic parent bodies is now widely accepted, and technological advancements allow for more precise and accurate measurements of amino acid concentrations. 
We collated the data from the following sources into a single large table. Figures \ref{CI} - \ref{CO} show concentrations of amino acids per meteorite, within each meteorite class. Concentrations are reported in parts-per-billion (ppb); for example, nanograms of amino acid per gram of meteorite. Where applicable, amino acid concentrations initially reported in left and right handed configuration abundances have been combined. 

\citet{Kaplan1963} was one of the first research groups to investigate the organic constituent present in stony meteorites. They analyzed the carbonaceous chondrite meteorites Orgueil, Cold Bokkeveld, Mighei, Murray, Felix, Lance, and Warrenton for the following proteinogenic amino acids: arginine, lysine, histidine, aspartic acid, glutamic acid, glycine, alanine, serine, proline, valine, threonine, leucine, tyrosine, phenylalanine, cysteine, and methionine.

The CM-type meteorite Murchison is one of the best studied meteorites of the carbonaceous chondrites. \citet{Cronin1983} quantified the abundances of the following amino acids: glycine, alanine, valine, proline, leucine, isoleucine, aspartic acid, and glutamic acid. In addition, they presented an amino acid abundance comparison of three different Murchison samples taken from the Field Museum of Natural History in Chicago, Illinois, Collections of Arizona State University, and the Smithsonian Institution National Museum in Washington D.C.

\citet{Peltzer1984} studied amino acids found in Murchison, exclusively those created via Strecker-type pathways. They discovered appreciable quantities of both glycine and alanine.

Around the same time, \citet{Shimoyama1985} investigated organics in Antarctic meteorites, looking specifically into amino acid concentrations on the Yamato-791198 meteorite.  They found measurable quantities of aspartic acid, threonine, serine, glutamic acid, proline, glycine, alanine, valine, isoleucine, and leucine.

\citet{Botta2002} present amino acid abundance data for a wide variety of CM, CI, CR, CV, and one ungrouped meteorite. Meteorites include CM-types Murchison, Murray, Nogoya, Mighei, and Essebi, CI-types Orgueil and Ivuna, CR-type Renazzo, CV-type Allende, and the ungrouped Tagish Lake meteorite. Proteinogenic amino acids detected were aspartic acid, serine, glutamic acid, glycine, and alanine.

\citet{Glavin2006} present total amino acid abundances in CM meteorites Murchison (to be identified as Murchison USNM 6650 in \citet{Glavin2011}), Lewis Cliffs (LEW) 90500, and Allan Hills (ALH) 83100. Amino acids reported include aspartic acid, glutamic acid, serine, glycine, alanine, and valine. \citet{Glavin2011} present an extensive account of total amino acid abundances in a variety of CI, CM, and CR class meteorites. Of the proteinogenic amino acids, the data include abundances for aspartic acid, glutamic acid, serine, glycine, alanine, and valine. Samples of CI meteorite Orgueil, CM meteorites Meteorite Hills (MET) 01070, Scott Glacier (SCO) 06043, Murchison USNM 5453, and Lonewolf Nunataks (LON) 94102, and CR meteorites Grosvenor Mountains (GRO) 95577, Elephant Moraine (EET) 92042, and Queen Alexandra Range (QUE) 99177 were all analyzed and reported in \citet{Glavin2011}. Also included in the \citet{Glavin2011} paper are the data for CM meteorites Murchison USNM 6650 and LEW 90500, though these 
samples were analyzed in \citet{Glavin2006}.

\citet{Martins2007a} perform chemical analyses of the organic content in the carbonaceous chondrite CR-type Shi\c{s}r 033. They discovered aspartic acid, glutamic acid, serine, glycine, valine, and leucine. \citet{Martins2007} published another paper in the same year. They investigated the organic content of three Antarctic CR meteorites: EET 92042, Graves Nunataks (GRA) 95229, and GRO 95577, and confirmed and quantified the presence of proteinogenic amino acids aspartic acid, glutamic acid, serine, glycine, alanine, and valine. 

\citet{PizzarelloHolmes2009} analyzed two CR meteorites, LaPaz Icefield (LAP) 02342 and GRA 95229. They reported abundance data for glycine, alanine, valine, isoleucine, leucine, aspartic acid, glutamic acid, serine, threonine, proline, phenylalanine, and tyrosine.

\citet{Monroe2011} analyzed the CM-type Bells meteorite and found aspartic acid, serine, glutamic acid, glycine, alanine, valine, leucine, and isoleucine.

\citet{Burton2012a} reported amino acid abundances for a wide variety of CV meteorites. Analyzed meteorites include CV3 types ALH 84028, EET 96026, LAP 02206, GRA 06101, Larkman Nunataks (LAR) 06317, and Allende, and CO3 types ALH A77307, Miller Cliffs (MIL) 05013, and Dominion Range (DOM) 08006. They report abundance data for the following proteinogenic amino acids: aspartic acid, glutamic acid, serine, glycine, alanine, and valine.

\citet{Chan2012} investigated the amino acid content of two CO meteorites, Colony and Ornans. They quantified the presence of proteinogenic amino acids alanine, valine, glycine, leucine, aspartic acid, and glutamic acid.

The reported amino acid abundances cited above, and used for our plots below, were obtained by the researchers by a combination of two widely practiced methods of amino acid extraction. These are high-performance liquid chromatography (HPLC) and a combination gas chromatography/mass spectroscopy (GC/MS). 
For a comparison of the methods of amino acid detection and quantification in meteorites, as well as a discussion of their limitations, see \citet{BottaBada}, \citet{Glavin2006}, and \citet{Martins2007}. Original sources for the data from each meteorite listed above are given in Table~\ref{sources}.

\begin{table*}[t]
\caption{Citations for Associated Meteorites}
\label{sources}
\begin{longtable}{l l l l}
Type & Meteorite & Sample Number, if Applicable & Reference(s) \\ \hline 
CI1 & Ivuna & N.A.& \citet{Botta2002} \\ 
CI1  & Orgueil & N.A. & \citet{Botta2002,Glavin2011} \\ 
CM1 & Meteorite Hills & MET 01070 & \citet{Glavin2011}\\
CM1& Scott Glacier & SCO 06043 & \citet{Glavin2011}\\
CM2 & Cold Bokkeveld & N.A.& \citet{Kaplan1963} \\
CM2 & Mighei & N.A.& \citet{Kaplan1963,Botta2002} \\
CM2 & Murray & N.A. & \citet{Kaplan1963,Botta2002} \\
CM2 & Yamato & Yamato 791198 & \citet{Shimoyama1985} \\
CM2 & Nogoya & N.A.& \citet{Botta2002} \\
CM2 & Essebi & N.A.& \citet{Botta2002} \\
CM2 & Allan Hills & ALH 83100 & \citet{Glavin2006} \\
CM2 & Murchison & N.A.& \citet{Botta2002} \\
CM2 & Murchison & Field Museum & \citet{Cronin1983} \\
CM2 & Murchison & AZ State Univ. & \citet{Cronin1983} \\
CM2 & Murchison & Smithsonian & \citet{Cronin1983} \\
CM2 & Murchison & USNM 5453 & \citet{Glavin2011} \\
CM2 & Murchison & USNM 6650 & \citet{Glavin2006} \\
CM2 & Lonewolf Nunataks & LON 94102 & \citet{Glavin2011} \\
CM2 & Lewis Cliffs & LEW 90500 & \citet{Glavin2006} \\ 
CM2 & Bells & N.A.& \citet{Monroe2011} \\
CR2 & Renazzo & N.A.& \citet{Botta2002} \\
CR1 & Grosvenor Mountains & GRO 95577 & \citet{Martins2007,Glavin2011} \\
CR2 & Elephant Moraine & EET 92042  & \citet{Martins2007,Glavin2011} \\
CR2 & Graves Nunataks & GRA 95229  & \citet{Martins2007,Pizzarello2009} \\
CR2 & LaPaz & LAP 02342  & \citet{Pizzarello2009}\\
CR2 & Shi\c{s}r & Shi\c{s}r 033 & \citet{Martins2007a}\\
CR2 & Queen Alexandra Range & QUE 99177  & \citet{Glavin2011}\\ 
CV3 & Allan Hills & ALH 84028  & \citet{Burton2012a} \\
CV3 & Elephant Moraine & EET 96026  & \citet{Burton2012a} \\
CV3 & LaPaz Icefield & LAP 02206  & \citet{Burton2012a} \\
CV3 & Larkman Nunatak & LAR 06317  & \citet{Burton2012a} \\
CV3 & Graves Nunataks & GRA 06101 & \citet{Burton2012a} \\
CV3 & Allende & N.A. & \citet{Botta2002,Burton2012a} \\ 
CO3 & Miller Range & MIL 05013 & \citet{Burton2012a} \\
CO3 & Dominion Range & DOM 08006 & \citet{Burton2012a} \\
CO3 & Allan Hills & ALHA 77307 & \citet{Burton2012a} \\ 
CO3 & Colony &  N.A.& \citet{Chan2012} \\
CO3 & Ornans &  N.A.& \citet{Chan2012} \\
\end{longtable}
\end{table*}

\section{Results: Amino Acid Abundances and Relative Frequencies}

 \begin{figure}
     \centering
      \includegraphics[width=87mm]{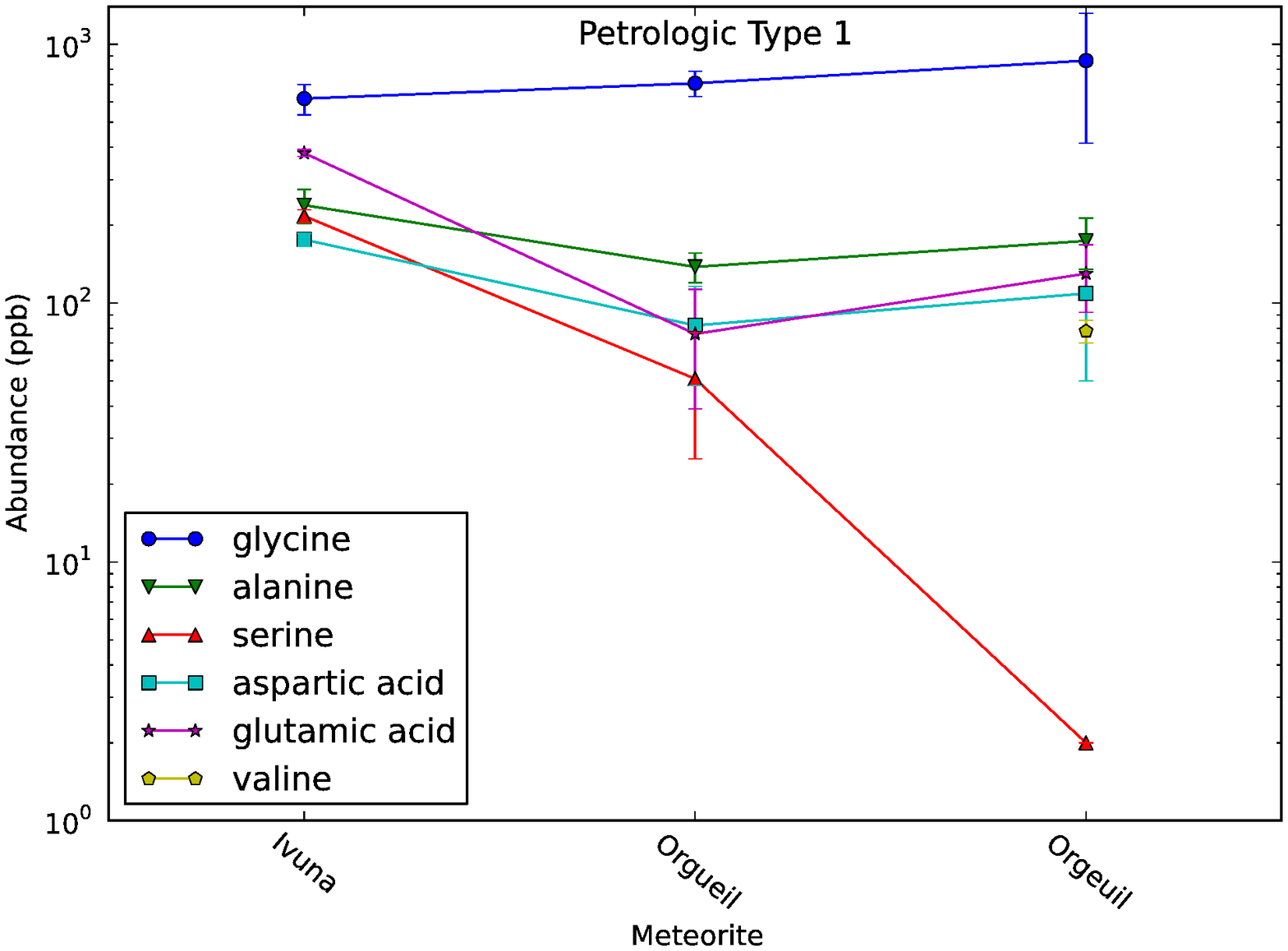}
      \caption{Amino acid abundances for three samples of CI meteorite. Concentrations are shown in ppb and range from 2 to 865.}
      \label{CI}
    \end{figure}
  \begin{figure*}
     \centering
      \includegraphics[width=\textwidth]{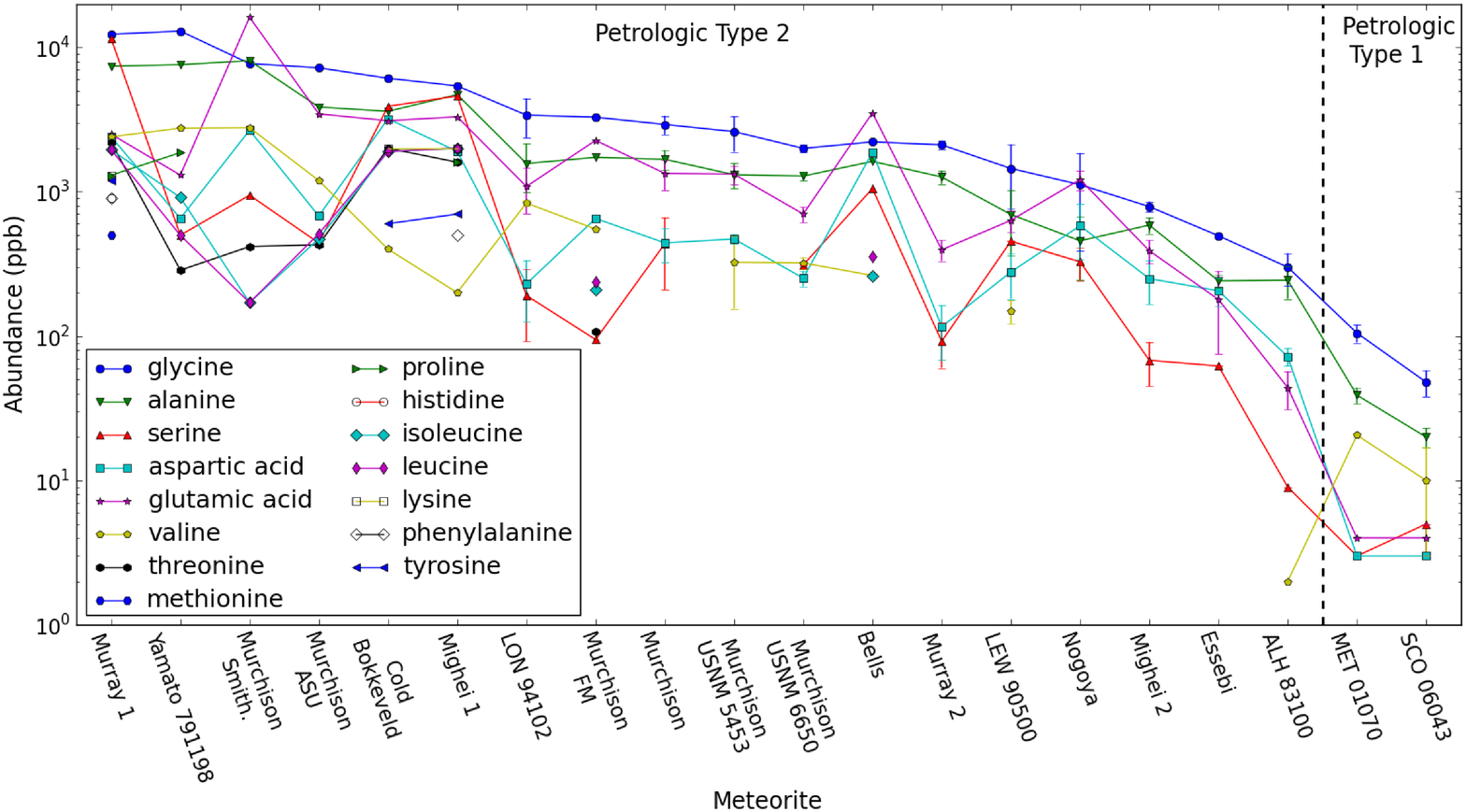}
      \caption{Amino acid abundances for 20 samples of CM meteorite. Concentrations are shown in ppb and range from 2 to 16000.}
      \label{CM}
    \end{figure*}     
    
\begin{figure}
     \centering
      \includegraphics[width=87mm]{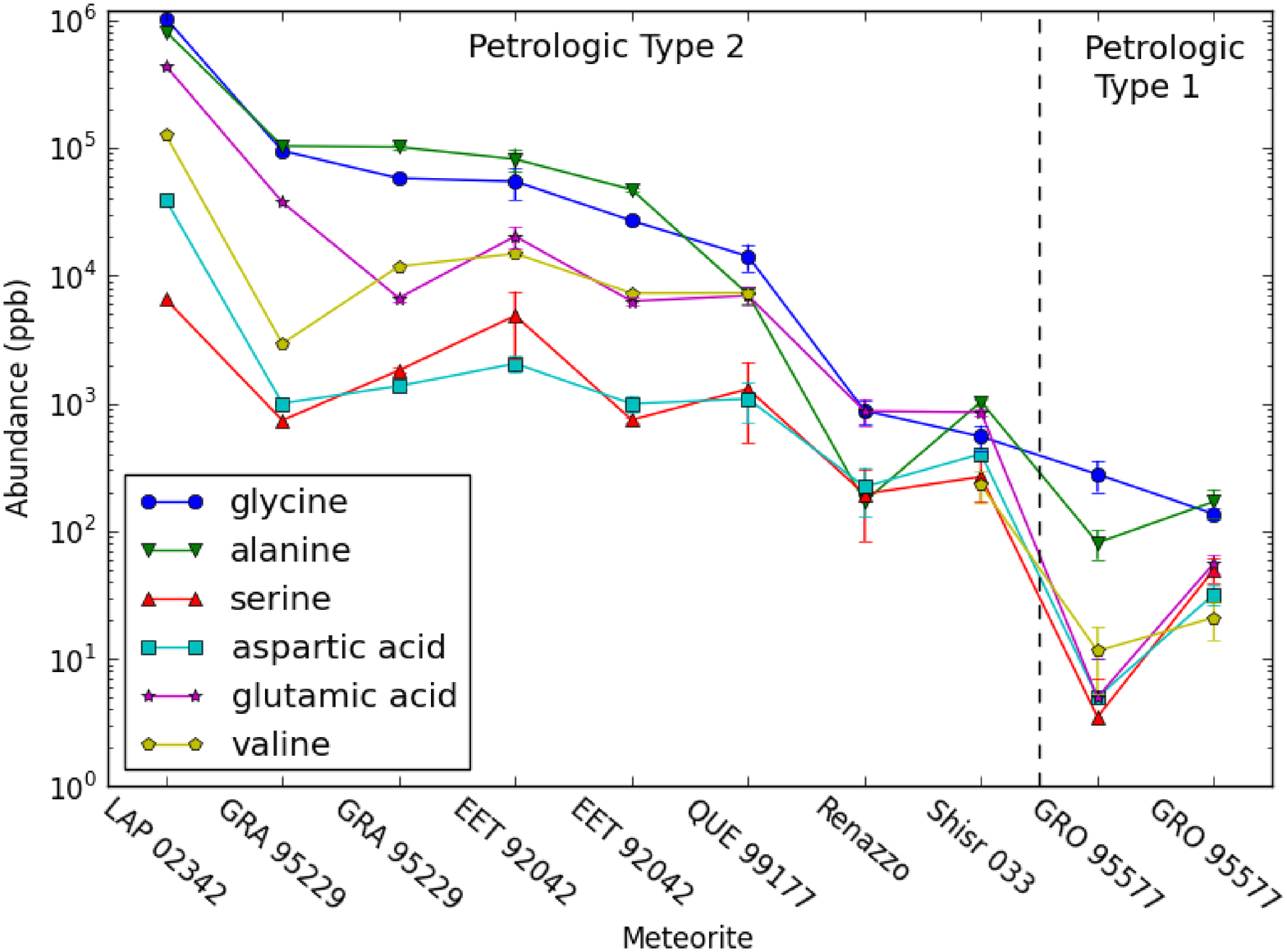}
      \caption{Amino acid abundances for 10 samples of CR meteorite. Concentrations are shown in ppb and range from 4 to 1032000.}
      \label{CR}
    \end{figure}

\begin{figure}[ht!]
     \centering
      \includegraphics[width=87mm]{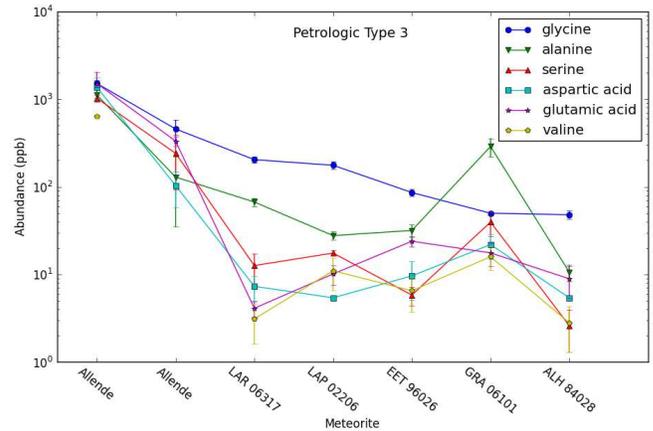}
      \caption{Amino acid abundances for seven samples of CV meteorite. Concentrations are shown in ppb and range from 2 to 1500.}
      \label{CV}
    \end{figure}

\begin{figure}[ht!]
     \centering
      \includegraphics[width=87mm]{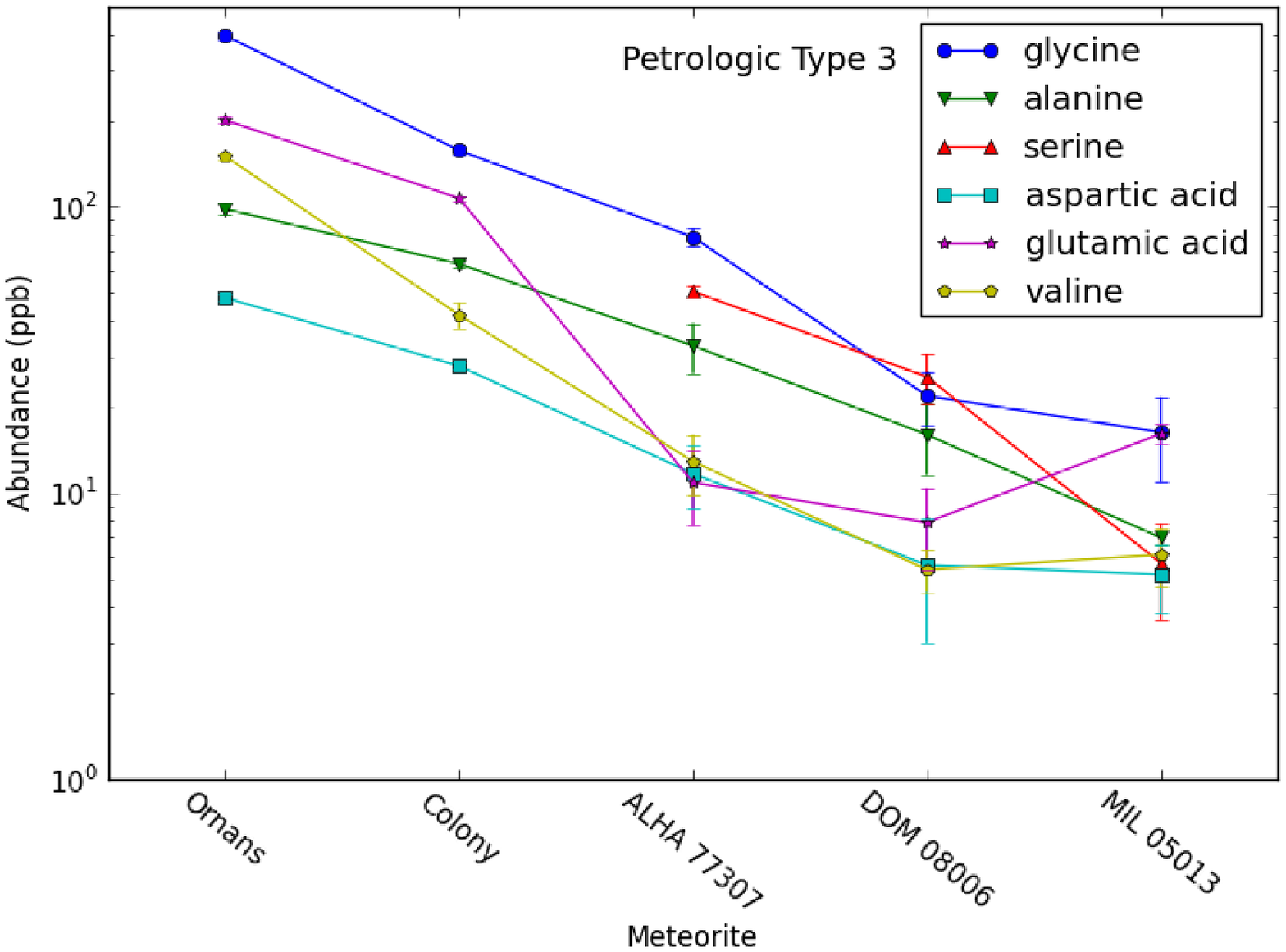}
      \caption{Amino acid abundances for five samples of CO meteorite. Concentrations are shown in ppb and range from 5 to 400.}
      \label{CO}
    \end{figure}

Using amino acid abundance data from the above sources, we created a series of figures (Figures \ref{CI} - \ref{CO}) showing amino acid abundances across a range of meteorites within each subclass. To visualize more easily the patterns in amino acid spread, the following figures show amino acid abundances on log scales, separated by meteorite class. The data are presented graphically to improve pattern recognition. We use line graphs to aid the eye in tracing the abundance of a given amino acid across a variety of meteorites. The following figures are ordered in a sequence outlined by the petrographic classification in Figure~\ref{petrology}. The ordering of meteorites within each figure (within each subclass) is arbitrary. However, we chose to order the sequence of meteorites in the data generally monotonically with their glycine abundance. Glycine is the simplest, most energetically favorable, and largely the most abundant amino acid. 

Figure \ref{CI} shows amino acid concentrations for the CI-type meteorites. There are three CI meteorites shown here, Ivuna and two samples of Orgueil (from separate analyses). All three are petrologic type 1. Glycine is the most abundant amino acid in each meteorite. Alanine, glutamic acid, aspartic acid, and valine remain relatively constant at a mid-level abundance. Serine remains near the bottom in all samples, as well, a pattern that seems to be consistent throughout the meteorite classes. There are no significant fluctuations in concentration for any amino acid. The CI subclass contains simple assortment of amino acids, primarily the amino acids glycine and alanine. The majority of CI abundance data lay between $\sim 10$ and $400$ ppb, with a few exceptions around $800$ ppb.


Concentrations for the CM subclass are shown in Figure \ref{CM}. There are 20 sets of data for CM meteorites: two samples of Murray, called Murray 1 and Murray 2, Yamato-791198, six samples of Murchison, called Murchison, Murchison Smith., Murchison ASU, Murchison FM, Murchison USNM 5453, and Murchison USNM 6650, Cold Bokkeveld, Mighei, Bells, LON 94102, LEW 90500, Nogoya, a second sample of Mighei, Essebi, ALH 83100, MET 01070, and SCO 06043. The first 18 are all type 2 meteorites. The last two, MET 01070 and SCO 06043, are type 1. As in Figure \ref{CI}, glycine again rides along the top of Figure \ref{CM}. Alanine follows in a close second for most abundant amino acid of the CMs. Serine remains near the bottom abundance levels in most meteorites. The wider variety of CM-types allows us a broader landscape to observe, which also allows for more variation in amino acid concentration. We see a considerable fluctuation in abundance of numerous amino acids other than glycine, especially around the Cold Bokkeveld and Mighei meteorites, and again at Bells. In general, we see that the amino acid tracks remain roughly parallel (minimal variation) across the CM class. The average concentration for a single amino acid among CM-types is on the order of $10^3$ ppb. Most concentrations range from $10^2$ to $10^4$ ppb. 

Figure \ref{CR} shows amino acid concentration for the CR subclass of meteorites. We have 10 meteorite samples for the CRs. One CR meteorite is petrologic type 1, GRO 95577, for which there are two data sets from two separate analyses. The other eight samples are type 2: Shi\c{s}r 033, Renazzo, QUE 99177, two samples of GRA 95229, two samples of EET 92042, and LAP 02342. Note that QUE 99177 has occasionally been classified as a CR3 type rather than a CR2, as in \citet{Glavin2011}. Amino acid abundances for this meteorite are between $10^3$ and $10^4$ ppb, well above the average abundance for other type 3 samples. For the purpose of this work, we consider QUE 99177 as a type 2 carbonaceous chondrite. Glycine continues largely to be the most abundant amino acid present, and serine among the least abundant. Valine, aspartic acid, and glutamic acid tracks move roughly parallel near the middle of the abundance range. CR-types show the greatest variation in abundance of any subclass; abundances span approximately 
five orders of magnitude. An average amino acid concentration among CR chondrites is approximately $10^4$ ppb. The full range of abundance in CR-types shows the greatest variation in concentration, from $\sim 10^1$ to $10^6$ ppb.

Figure \ref{CV} shows concentration data for the CV meteorites. There are seven sets of amino acid abundance data for the CV class, all CV3-types. These chondrites are Allende (data from two different analyses), LAR 06317, EET 96026, LAP 02206, GRA 06101, and ALH 84028. Again glycine predominantly leads the abundance data curves. Valine and aspartic acid seem to be the least abundant amino acids amongst the CV-types. Serine, alanine, and glutamic acid trace near the middle. There is a single large spike in alanine values seen in the GRA 06101 meteorite. CV-types average between $10$ and $200$ ppb, with a single meteorite specimen showing individual amino acid abundances around $10^3$ ppb. 

Amino acid concentrations within the CO-type meteorites are shown in Figure \ref{CO}. We have data for five CO3 chondrites: Ornans, Colony, ALHA 77307, DOM 08006, and MIL 05013. Figure \ref{CO} amino acid concentration tracks are among the most parallel of any meteorite class. Glycine consistently remains one of the most abundant amino acids, among the CO-types and all other meteorite classifications. In this type, valine and aspartic acid are the least abundant. Alanine, glutamic acid, and serine show some variation. Amino acids are the least abundant in CO-type meteorites. Abundance levels across the CO subclass range between $5$ and $400$ ppb.

Overall patterns that emerge include the tendency of glycine to be the most abundant amino acid, and serine, valine, and aspartic acid tend to be the least abundant. Alanine and glutamic acid show significant variation across all classes. In general, amino acid abundance curves track roughly parallel across all meteorites within a given subclass. 

To the best of our knowledge, we are the first to visualize the data in this way. The collective data pool of amino acid abundances in meteorites is vast. In these figures meteorites are organized according to petrographic type, allowing us to identify patterns in abundance with changes in chemical composition and temperature. Abundance tracks are viewed relative to glycine, allowing for identification of irregularities in both abundance and frequency, which we will model in Paper II.

\subsection{Amino Acid Abundance Patterns}

\begin{figure*}
     \centering
      \includegraphics[width=.95\textwidth]{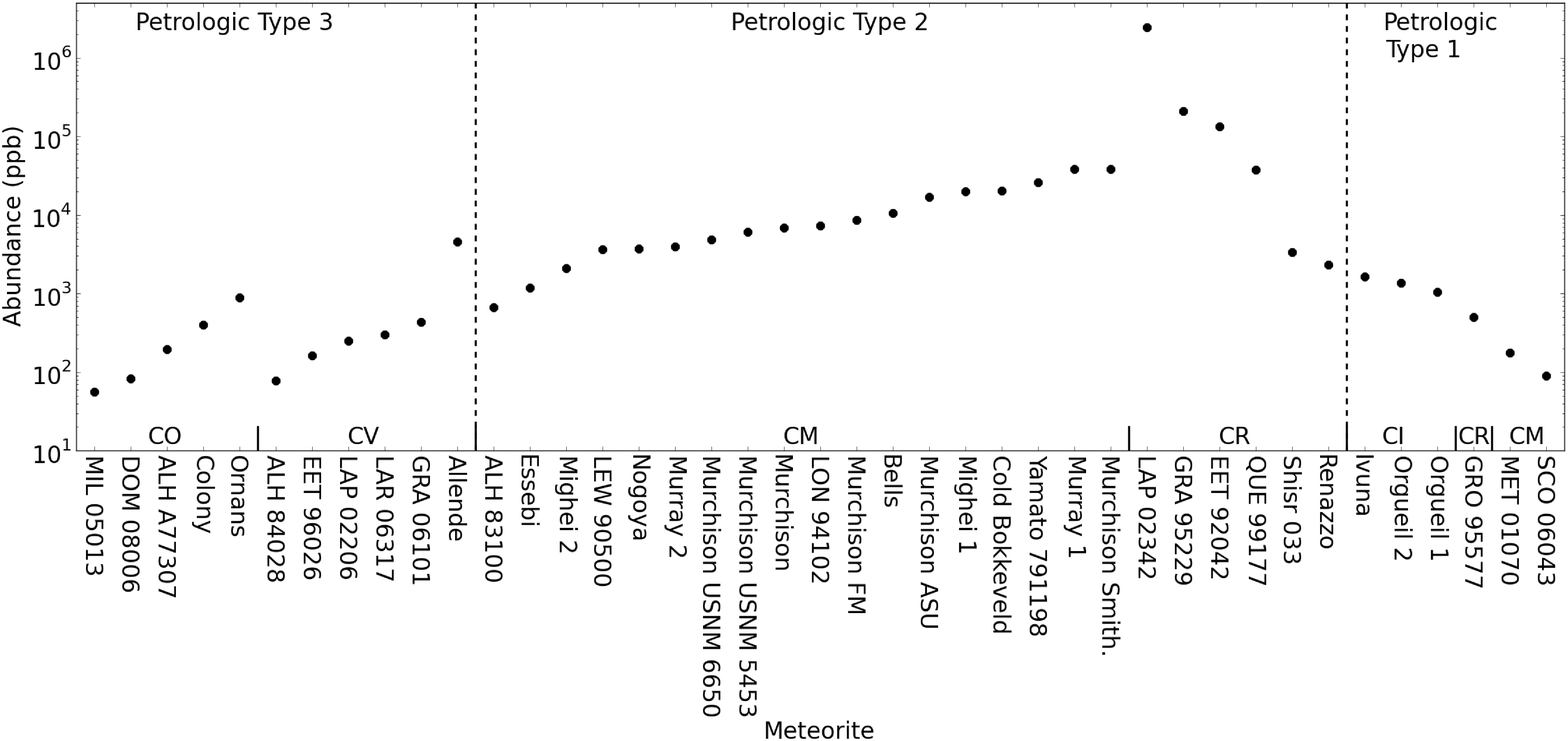}
      \caption{Total amino acid abundances for all meteorites. Summed amino acid values include glycine, alanine, serine, aspartic acid, glutamic acid, and valine. Concentrations are shown in ppb.}
      \label{total}
    \end{figure*}

To observe the overall trends total amino acid abundance, we plot in Figure \ref{total} the total amino acid concentrations for the entire range of meteorites across all classifications. Each meteorite is represented by a single data point that is the sum of amino acid abundances for six amino acids: glycine, alanine, serine, aspartic acid, glutamic acid, and valine. We have the most complete set of data for these six amino acids. In the cases of multiple sets of abundance data for the same meteorite, amino acid concentrations were averaged. We did not include incomplete sets of abundance data when calculating the sum for each meteorite, as inclusion of extra amino acids for some meteorites and not for others would skew the total abundance pattern.

    \begin{figure*}
     \centering
      \includegraphics[width=.95\textwidth]{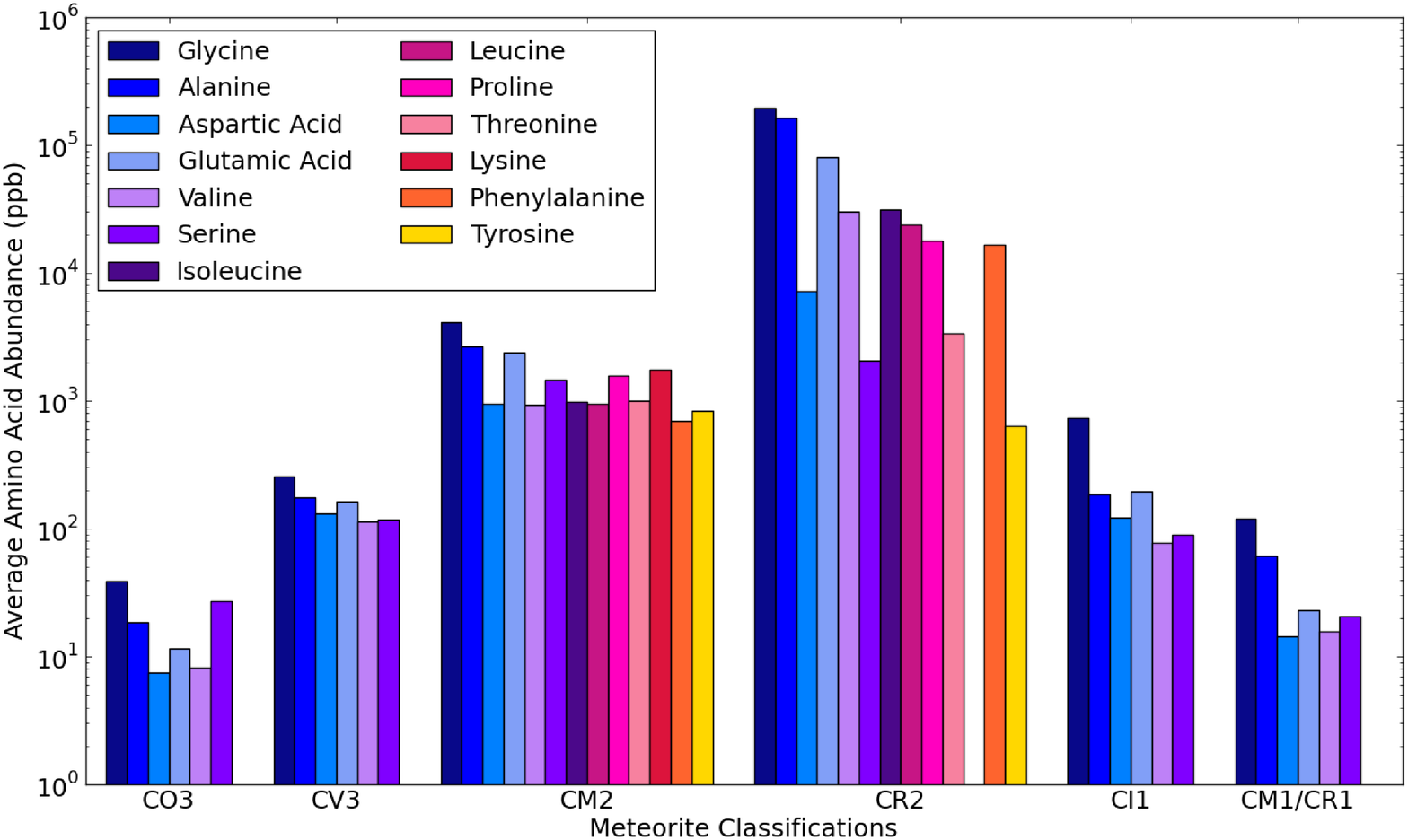}
      \caption{Average amino acid abundances separated by meteorite class. There are three meteorite samples of CI1, two samples of CM1, one of type CR1, 18 of CM2, six of CR2, six meteorites of type CV3, and three samples of CO3. The averages for the CM1 and CR1 classes are combined as there are so few samples of each type. CI1, CM1/CR1, CV3, and CO3 only have available and complete data for six amino acids: glycine, alanine, aspartic acid, glutamic acid, valine, and serine. The greater number of samples in the CM2 and CR2 classes allows for a broader spectrum of available amino acid data: glycine, alanine, aspartic acid, glutamic acid, valine, serine, isoleucine, leucine, proline, threonine, lysine, phenylalanine, and tyrosine.}
      \label{totalavg}
    \end{figure*}    

Within the broader spectrum of amino acids in meteorites, the data show general abundance trends among meteorite classes. We see that amino acids are most abundant in the CM and CR meteorite classes, as noted previously by, for example,  \citet{Martins2007}, \citet{Glavin2011}, and \citet{Burton2012a}. The average concentration for a single meteorite among CM2-types is on the order of $10^4$ ppb. Most total concentrations range from $10^3$ to $10^5$ ppb. Average total concentrations among CR2 chondrites are around $10^5$ ppb. The full abundance range in CR-types shows the greatest variation in concentration, from $10^3$ to $10^6$ ppb.  
Petrologic type 1 meteorites, which include the CI chondrites as well as a few CM/CR samples, have total abundances on the order of $10^2$ ppb. Petrologic type 3 chondrites, both CV- and CO-types, have total abundances on the same order of magnitude, near  $10^2$ ppb.

Figure \ref{totalavg} shows the average concentration of each amino acid per meteorite classification. We plot the amino acid abundance following the basic petrographic sequence - type 3, type 2, and type 1 - from least to most altered. The ordering of amino acids follows the sequence suggested by \citet{Higgs2009}. This figure represents one of the most important findings in this paper. There are fewer samples of meteorite types CI1, CM1/CR1, CV3, and CO3, where we have only a complete set of data for six amino acids. There is a larger set of meteorite samples in the CM2 and CR2 classes, and a broader variety of amino acids for which we have a complete set of data. 

As suggested by Figure \ref{petrology}, the CO and CV meteorites, all type 3, show relatively little aqueous alteration and associated low abundances of amino acids. The CM chondrites express a great variety of amino acids, including glycine, alanine, aspartic acid, glutamic acid, valine, serine, isoleucine, leucine, proline, threonine, lysine, phenylalanine, and tyrosine. The CR-types show a similarly broad variety in amino acid species; the same variety as the CM-types, with the exception of lysine. 
We see that the pattern of amino acid abundance roughly follows the classification according to petrologic type. The greatest total abundance of amino acids occurs in CM and CR type 2 meteorites. CI meteorites, all petrologic type 1, do contain a significant quantity of amino acids, though less than the CM2/CR2 by several orders of magnitude. The few meteorites we have of type CM1/CR1 show a similar quantity of total amino acids to CI1, indicating a dependence on formation processes associated with the type 1 petrologic type.  

Following the pattern laid out in Figure \ref{petrology}, we see an increase in amino acid synthesis associated with greater degrees of aqueous alteration and petrographic type 2, corresponding roughly to temperatures $0^{\circ}$C and $150^{\circ}$C. Higher temperatures, $\sim 150^{\circ}$C, correspond to the greatest degrees of aqueous alteration, amongst petrologic type 1 meteorites, though we see comparatively low abundances of amino acids in the type 1 samples compared to the type 2 meteorites. \citet{Martins2007} and \citet{Glavin2011} suggest this lack of significant amino acid presence may be due to some chemical oxidation process during aqueous alteration. 
Whatever the chemical reasons behind the decline in amino acid concentration, the data suggest that the observed peak in amino acid synthesis we see is due to the existence of optimal interior conditions inside meteoritic parent bodies, and temperatures associated with petrographic type 2 chondrites. 

The onion shell model for parent bodies, as in \citet{Weiss2013}, provides another possible explanation for the observed lack in amino acids among the more aqueously altered specimens of petrographic type 1. 
It is possible the material from which type 1 chondrites originate is formed from a section of the parent planetesimal that contains less water and/or organic material. \citet{Young1999} and \citet[and references therein]{Weiss2013} discuss additional distinguishing characteristics of the CI meteorites. The CI-types have oxygen isotope compositions as well as different bulk densities that are noticeably different than those of the other carbonaceous chondrite subclasses.



\subsection{Normalized Amino Acid Relative Frequencies}

\begin{figure}[h!]
     \centering
      \includegraphics[width=80mm]{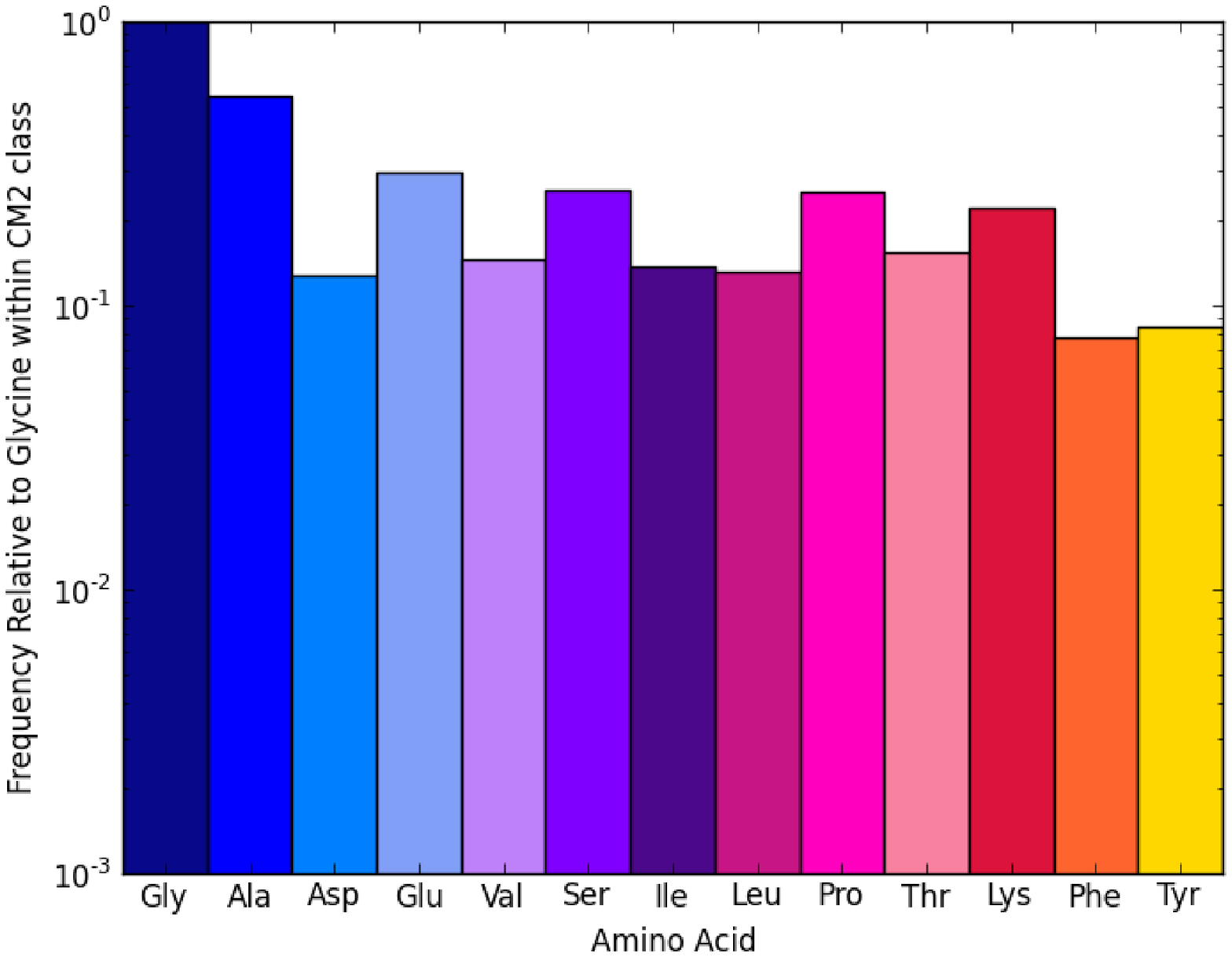}
      \caption{Relative amino acid frequencies for the CM2 meteorite class. Abundance values have been averaged over the 18 meteorite samples in this classification. Abundances of 13 amino acids are shown relative to glycine, reported in moles amino acid per moles glycine.}
      \label{CMrel}
    \end{figure}

The data visualized in the previous figures show that there exists a greater abundance of amino acids in meteorites of type CM2 and CR2. Given that standing pattern, we now take a closer look at these two subclasses in particular. Figure~\ref{CMrel} shows average amino acid frequencies relative to glycine for meteorites in the CM2 class. We normalize all frequencies to glycine, since it is the simplest and one of the most abundant amino acids, as well as the most energetically favorable. The ordering of amino acids follows the pattern in \citet{Higgs2009} of the 10 early-type amino acids. \citet{Higgs2009} found a thermodynamic relationship between Gibbs free energies of reaction and the observed rank in abundance of these 10 amino acids.

\begin{figure}[h!]
     \centering
      \includegraphics[width=80mm]{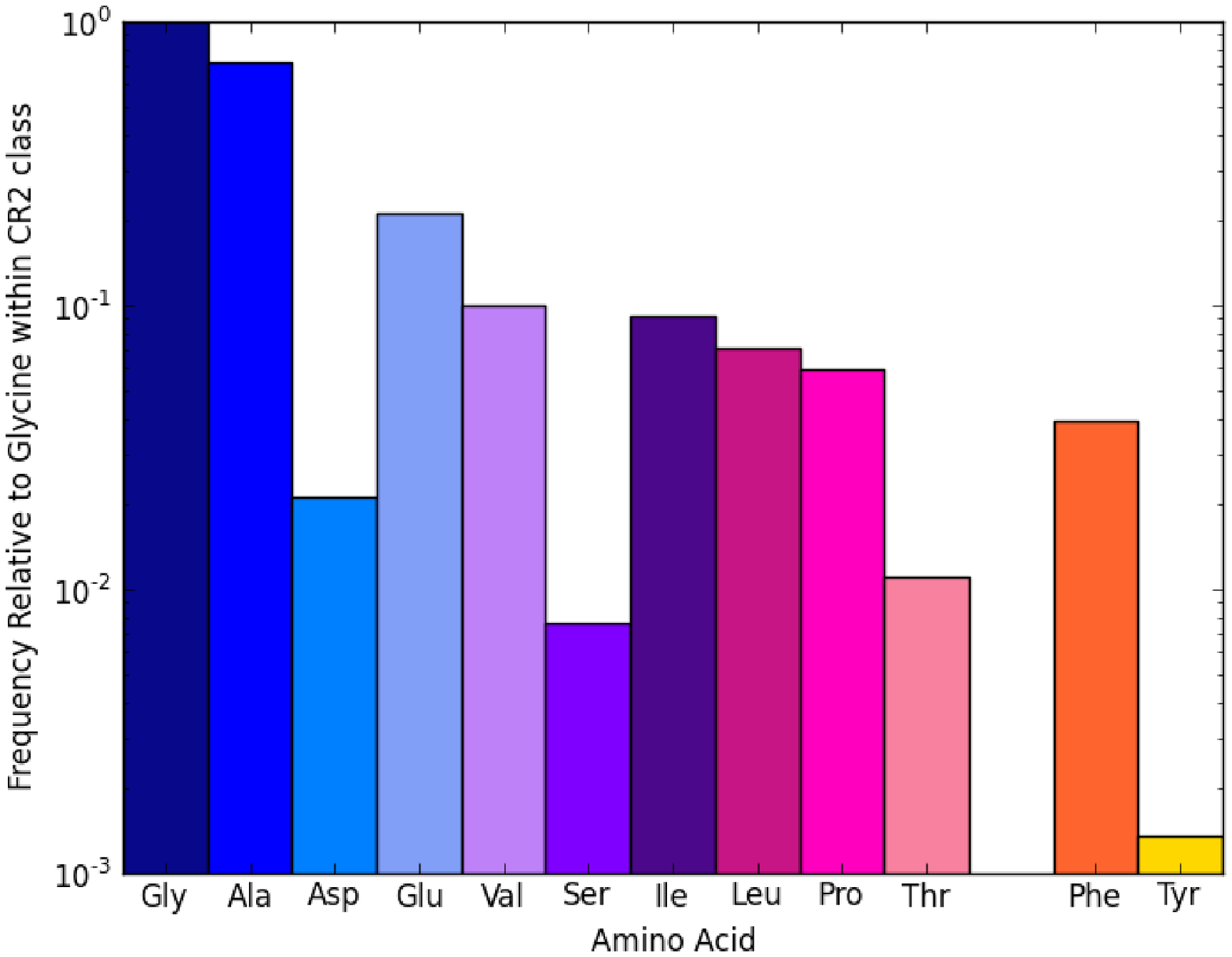}
      \caption{Relative amino acid frequencies for the CR2 meteorite class. Abundance values have been averaged over the 6 meteorite samples in this classification. Abundances of 12 amino acids are shown relative to glycine, reported in moles amino acid per moles glycine.}
      \label{CRrel}
    \end{figure}

Figure~\ref{CRrel} shows average amino acid frequencies relative to glycine for meteorites in the CR2 class. Note that the normalization to glycine does not include normalization over molecular weight. 
In both Figures \ref{CMrel} and \ref{CRrel} we see a relative predominance of glycine. Glycine is classically one of the most abundant amino acids found on meteorites, it is achiral, and is the easiest to synthesize. Alanine is the second most abundant amino acid in meteorites, followed by glutamic acid. In CM2-types, there are also appreciable quantities of serine, proline, and lysine. Other amino acids present in much smaller quantities are aspartic acid, valine, threonine, isoleucine, leucine, phenylalanine, and tyrosine. 

These plots allow us to make comparative statements about relative concentrations within a given meteorite class. \citet{Higgs2009} found that amino acid frequency data for three CM2 meteorites (Murchison, Murray, and Yamato) could be correlated with the Gibbs free energy of reaction using a purely thermodynamic relation. In Paper II, we use the frequencies shown in Figures \ref{CMrel} and \ref{CRrel} as a baseline for our modeling software to examine a much broader class of data using predictive thermodynamic theory.

\section{Discussion and Conclusions}
This work investigates amino acid abundance patterns and relative frequencies in $\alpha$-proteinogenic amino acids of carbonaceous chondrites. We collated an extensive data set from the literature from a variety of sources covering a range of meteorites, including CI, CM, CR, CV, and CO, and petrologic types 1 - 3. Data have been organized so as to emphasize possible physical variables that control amino acid synthesis and abundance patterns. The data represent a wide variety of meteoritic parent body alteration states. Collation of this data allows for identification of patterns and the relationship between amino acid abundance and meteorite classification/petrologic type. We included amino acid abundances for 41 different samples of meteorite, covering five different subclasses of meteorite and three petrologic types. 

It is possible to view our results on amino acid abundance patterns in the context of an onion shell model. We may view the sequence in types from 3 to 1 as descending from the cooler surface of a planetesimal, which has not been significantly aqueously altered, to a deeper layer, type 2, which has temperatures and conditions ideally suited to amino acid synthesis, from which the CM2- and CR2-types originate. With further descent toward the hot core, one finds conditions sampled by the type 1 chondrites, where temperatures are perhaps too hot to support amino acid presence. Detailed 3D simulations of planetesimals  by \citet{Travis2005} support such a picture. Key results from our work are listed below.

\begin{enumerate}
 \item We present the data according to physical properties and trends in formation mechanisms. Type 1 meteorites, be they CI, CM, or CR, are the most aqueously altered specimens. Aqueous alteration during formation occurred at higher temperatures, possibly $50^{\circ}$C to $150^{\circ}$C, yet we do not observe a peak in total amino acid concentration at this location. All CI meteorites are petrologic type 1. There are only three listed type 1 CM/CR chondrites (two CM1 meteorites, one CR1 meteorite).
  \item The type 2 meteorites (mostly CM-types and some CR-types) are generally associated with cooler alteration temperatures. These range broadly from $0^{\circ}$C to $100^{\circ}$C (based on splitting the CR temperature range into its petrographic types), show the greatest amino acid concentrations of all petrologic groups. The CM2 and CR2 chondrites contain the greatest quantities and variety of amino acids of all subclasses. 
\item All CV and CO chondrites are type 3, associated with cool formation temperatures, roughly $0^{\circ}$C to $50^{\circ}$C. Type 3 meteorites are relatively unaltered from the time of their original formation. They have undergone minimal alteration, either aqueous or thermal metamorphism, and show a corresponding lack of significant amino acid presence.
\item These data indicate the existence of an optimal temperature range for the abundance of amino acids in planetesimals. The temperature range associated with type 2 chondrites, possibly ranging $0^{\circ}$C to $100^{\circ}$C, corresponds with amino acid abundances greater by several orders of magnitude.
\item For this optimal range, we identify the relative frequencies of amino acids relative to glycine for the CM2 and CR2 carbonaceous chondrites. In \citet{Higgs2009}, relative frequencies are shown to occur according to thermodynamic principles - the more complex the molecule, the greater the Gibbs free energy necessary to synthesize it under certain conditions. The more extensive data shown here reinforces the conclusion that an underlying thermodynamic argument may be responsible. 
\end{enumerate}

In A. K. Cobb \& R. E. Pudritz (2014, in preparation), we investigate computationally if temperature and aqueous conditions in a planetesimal account for the overall abundance patterns we see in the data as well as for relative frequencies within subclasses. 
\newline

We would like to thank Daniel Glavin, Stephen Freeland, Boz Wing, Aaron Burton, Jamie Elsila, Jason Hein, and Darren Fernandes for stimulating discussions. We also thank an anonymous referee for several helpful suggestions that significantly improved the paper. A.K.C. would like to thank the Canadian Astrobiology Training Program who supported her with a CATP Graduate Fellowship, funded by NSERC CRSNG. R.E.P. was supported by an NSERC Discovery Grant.

\newpage

\end{document}